\documentclass[aps,showpacs,twocolumn,prl,toolkits]{revtex4}

\usepackage{axodraw}
\usepackage{mathrsfs}
\usepackage{amsmath}
\usepackage{graphicx}
\usepackage{epstopdf}

\begin{document}

\title{Doubly heavy hadrons and the domain of validity of doubly heavy
diquark--anti-quark symmetry}

\author{Thomas D. Cohen}
\email{cohen@physics.umd.edu}

\affiliation{Department of Physics, University of Maryland,
College Park, MD 20742-4111}

\author{Paul M. Hohler}
\email{pmhohler@physics.umd.edu}

\affiliation{Department of Physics, University of Maryland,
College Park, MD 20742-4111}

\begin{abstract}
In the limit of heavy quark masses going to infinity, a symmetry
is known to emerge in QCD relating properties of hadrons with two
heavy quarks to analogous states with one heavy anti-quark. A key
question is whether the charm mass is heavy enough so that this
symmetry is manifest in at least an approximate manner.  The
issue is crucial in attempting to understand the recent reports
by the SELEX Collaboration of doubly charmed baryons.  We argue
on very general grounds that the charm quark mass is
substantially too light for the symmetry to emerge automatically
via colour coulombic interactions. However, the symmetry could
emerge approximately depending on the dynamical details of the
non-perturbative physics. To treat the problem systematically, a
new expansion that simultaneously incorporates NRQCD and HQET is
needed.

\end{abstract}


\maketitle

\section{Introduction}

It has been known for some time that in the limit of arbitrarily
large heavy quark masses that QCD has a symmetry which relates
hadrons with two heavy quarks (anti-quarks) to analogous states
with one heavy anti-quark (quark) \cite{savage}.  We will refer to
this symmetry as the doubly heavy diquark--antiquark (DHDA)
symmetry. Presumably when the masses are finite, but very large,
a remnant of this DHDA symmetry will survive in the form of an
approximate symmetry. A key issue is how large must the masses be
before such an approximate DHDA symmetry is manifest in a useful
way.  The issue is particularly relevant for charm quarks---both
because the charm quark is the lightest of the heavy quarks and
hence the approximation is most likely to fail and because doubly
bottomed hadrons (or hadrons with a charm and a bottom) are
presumably more difficult to create and detect than doubly
charmed ones.

\begin{figure}[htb]
\includegraphics[scale=.5]{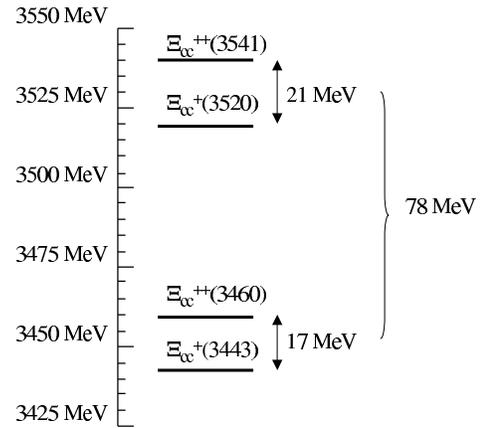}
  \caption{\label{fig:specxi} Spectrum of $\Xi_{cc}$ that have been observed by the SELEX Collaboration \cite{selex}.}
\end{figure}

The issue remained of only marginal importance in the absence of
observed doubly heavy hadrons.  However, in the past several
years, the SELEX Collaboration has reported the first sighting of
doubly charmed baryons \cite{selex}.  Four states, $\Xi_{cc}^+
(3443)$, $\Xi_{cc}^{++} (3460)$, $\Xi_{cc}^+ (3520)$, and
$\Xi_{cc}^{++} (3541)$ (which have been interpreted as two pairs
of iso-doublets) are reported, as shown in Fig.~\ref{fig:specxi}.
It should be noted that all four states were identified through
their weak decay products.  This is surprising as one would
ordinarily expect the excited states to decay electro-magnetically
much more rapidly and thus wash out a signal for weak decays.
This issue creates a potential problem for any interpretation of
the data. Additionally, most recently, BaBar has reported that
they have not observed any evidence of doubly charmed baryons in
$e^+ e^-$ annihilations \cite{babar}. However, we would set these
issues aside and take the existence of all four states as given
to ask whether the properties of these states could be understood
at least qualitatively in terms of the DHDA symmetry. Recently
refs.~\cite{bvr} and \cite{fleming} argued that the splitting
between the lower doublet and the upper doublet $\Xi$ states can
be understood semi-quantitatively (at the 30\% level) in terms of
an approximate DHDA symmetry.

This paper critically examines the extent to which an approximate
DHDA symmetry could be present for charm quarks. This is of
importance both for the doubly charmed states found by SELEX and
also for the existence of putative doubly charmed tetraquarks
which are known to exist in the heavy quark limit \cite{manohar}
and in potential models \cite{richard}. We find strong evidence to
suggest that the charm quark mass is {\it not} heavy enough for
the symmetry to emerge automatically of color coulombic
interactions. The key issue is the degree to which scales that
separate in the heavy quark limit (and whose separations are
critical to the derivation of the DHDA symmetry) in fact separate
for doubly charmed systems. As we will detail below, such a scale
separation probably does not hold. Despite this, we will show
that the presence of certain non-perturbative interactions could
result in an approximate DHDA symmetry in the charm sector.

To begin the discussion, let us consider why one expects the DHDA
symmetry. Physically, it arises from a diquark pair forming a
tightly bound nearly point-like object. The  attraction  between
the two heavy quarks in the diquark comes from a color coulombic
interaction that is attractive in the color $\boldsymbol{\bar{3}}$
channel. If the mass of the quarks is large enough, the heavy
quarks move slowly and act like non-relativistic particles in a
coulombic potential.  As the size of a coulombic bound state is
inversely proportional to its mass (for fixed coupling),  in the
large mass limit the diquark becomes a heavy, small object with
color $\boldsymbol{\bar{3}}$.  To a good approximation it becomes
a static point-like $\boldsymbol{\bar{3}}$ color source; in this
sense it acts in essentially the same way as a heavy anti-quark.
This symmetry was first discussed by Savage and Wise \cite{savage}
in the context of relating the properties of doubly heavy baryons,
$QQq$, to those of heavy mesons, $\bar{Q}q$.

To the extent that one can treat the heavy diquark as formed, one
can simply use standard heavy quark effective theory (HQET) to
describe the properties of the doubly heavy baryons.  Since the
diquark in the doubly heavy baryon essentially acts as an
antiquark, one can directly relate the properties of this system
to heavy mesons. Using the HQET effective Lagrangain in
ref.~\cite{savage}, a relationship valid at large $M_Q$ for the
mass difference of spin excited states between the doubly heavy
baryons and heavy mesons was derived \footnote{Equation
(\ref{eq:sw}) is different from that in ref.~\cite{savage} by a
factor of $\frac{1}{2}$.  This error was observed and corrected
by ref.~\cite{ebert} and \cite{bvr}.}:
\begin{equation}\label{eq:sw}
m_{\Sigma^{*}}-m_{\Sigma} = \frac{3}{4} (m_{P^{*}}-m_{P}),
\end{equation}
where $\Sigma$ and $\Sigma^{*}$ are the doubly heavy anti-baryons
with $S=\frac{1}{2}$ and $S=\frac{3}{2}$, respectively,  and $P$
and $P^{*}$ are the heavy mesons with $S=0$ and $S=1$,
respectively.  From the prospective of HQET, this relationship
should hold to $O(\Lambda_H^2/m_Q)$ where $\Lambda_H$ is a typical
hadronic energy and is proportional to, but not identical to
$\Lambda_{QCD}$. However as was discussed in ref.~\cite{bvr}, the
finite size of the diquark gives rise to corrections formally
larger than this in the large mass regime. At the time of the
Savage and Wise paper, this relationship was a prediction of the
theory: doubly heavy baryons had not been discovered. The SELEX
data will allow us to explore this relation with some real world
data.

Before proceeding further, we should note that this analysis is
based on the assumption that a spatially small and tightly bound
diquark configuration exists and remains unexcited in the
dynamics. The key question we address is the extent to which this
assumption is true.

To examine the issue of diquark excitations, a systematic
treatment for the dynamics of two heavy quarks is needed.  At a
formal level the non-relativistic expansion of the heavy quark
degrees of freedom with QCD (NRQCD) is the natural language to
explore this issue. NRQCD was first developed by Bodwin, Braaten,
and Lepage \cite{BBL}, where it was modeled after a similar
treatment in the context of QED \cite{cl}. HQET is generally
considered as an expansion in powers of $p/m$, with $p \sim
\Lambda_{H}$. Thereby it creates two energy scales, $m$ and
$\Lambda_{H}$. On the other hand, NRQCD requires the introduction
of two new scales: the characteristic momentum, $mv$, and energy
scale, $mv^2$, where $v \sim \alpha_s(mv)$ is the characteristic
velocity of the two heavy quarks relative to each other. With the
hierarchy, $m \gg mv \gg mv^2$, the characteristic regimes in
terms of (energy, momentum) of the heavy quarks are: ($m$, $m$),
($mv$, $mv$), ($mv^2$, $mv$), and ($mv^2$, $mv^2$). These are
conventionally referred to as hard, soft, potential, and
ultrasoft, respectively. Traditional NRQCD has been further
simplified into two different effective theories, pNRQCD and
vNRQCD.  pNRQCD integrates out the soft momentum gluons to form
heavy diquarks states  with definite color, and uses these diquark
states as the degrees of freedom \cite{brambilla}. On the other
hand, vNRQCD keeps the heavy quarks as explicit degrees of freedom
while matching the effective theory at the hard scale
\cite{luke}. In all forms of NRQCD, the separation of scales
creates an expansion of powers of $v$.

On physical grounds, one expects that the NRQCD at leading order
of systems with two heavy quarks (or anti-quarks) ought to reduce
to the HQET description of the dual problem---{\it i.e.}, the
problem related by the DHDA symmetry. Recently, ref.~\cite{bvr}
derived the presence of DHDA symmetry in the context of pNRQCD
while ref.~\cite{fleming} confirmed this for vNRQCD by showing
the equivalence between vNRQCD and pNRQCD. It should be noted
that this derivation represents a qualitatively new domain for
NRQCD. Traditionally, NRQCD is applied to systems with one heavy
quark and one heavy anti-quark with no valance light quark
degrees of freedom. The fact that the technique may be extended
to problems with two heavy quarks plus additional light quark
degrees of freedom is non-trivial. One central point, that should
be stressed, is that the derivation is quite general and applies
equally well to the problem of heavy tetraquarks as well as
doubly heavy baryons. The key advantage to the NRQCD formalism is
that corrections to this symmetry can be systematically
incorporated by working at higher order.

While it is known that the DHDA symmetry must emerge in the heavy
quark limit, it is not immediately clear how large the corrections
to the symmetry results should be for the realistic case in which
heavy quarks have large but finite mass. Clearly the fundamental
issue is the interplay between the diquark binding into an
approximately point-like object and the extent that the diquark
is point-like from the light quarks perspective; thus both the
details of the physics of the interactions between the two heavy
quarks as well as the between the heavy and light quarks are
essential. Previous work in this area, \cite{bvr,fleming}, have
concentrated their efforts on perturbative expansions of the
interactions between the two heavy quarks in the framework of
NRQCD, and have not dealt with heavy/light interactions. Since the
interactions between the heavy and light quarks are intrinsically
non-perturbative, it cannot be estimated directly via the
techniques of NRQCD. The full expansion should be a combination
of HQET and NRQCD that incorporates the mixing of perturbative and
non-perturbative scales. The issue of how to attack the question
of the scale of these corrections for charmed or bottom quarks is
the motivation for this paper. We do this in the context of the
SELEX data with tools motivated by NRQCD. Even though we do not
fully formulate the new combined expansion in this paper, we
provide strong arguments suggesting the need for such a theory
when dealing with doubly heavy mesons.  This paper explores this
issue both in terms of systematic treatment of the problem based
on power counting in effective field theories and in terms of more
heuristic phenomenological reasoning.

We divide this paper into two major sections.  In the first, we
work in the large quark mass limit, and develop the consequences
of the spectrum in this regime.  In the second section, we work
with a finite quark mass and present arguments that show the
SELEX data is not consistent with the large mass limit, the need
for a new expansion to describe this system, and the justification
beyond NRQCD of the apparent DHDA symmetry seen by SELEX.

\section{Consequences of DHDA symmetry in the large mass limit}

Before addressing the key question of whether the charm quarks
are too light for the DHDA symmetry to be manifest, it is useful
to consider just what implications the DHDA symmetry has on the
spectrum when the symmetry is manifest---namely, when the quarks
are sufficiently heavy. We attempt to consider the extreme limit,
where all relevant scales cleanly separate. It is unlikely that
the physical world exists in this limit. Nevertheless, an
understanding of the the physics in this extreme regime is useful
in understanding the applicable expansions. There has been
extensive work using a variety of models in detailing the hadronic
spectrum including \cite{richard2} and \cite{kiselev} among
others. Our focus here will be considering the spectrum in the
context of a possible DHDA symmetry. We will consider a more
modest regime, that is intended to describe the physical world,
in the next section.

The first consequence we consider is qualitative---namely, the
existence of exotic states. The DHDA symmetry in HQET was first
used to relate doubly heavy baryons to heavy mesons \cite{savage}.
However, the symmetry is independent of the light quarks in the
problem.  Formally, in NRQCD, the light quarks are govern by
non-perturbative dynamics, and are thereby considered irrelevant
when focusing on the heavy quarks in the large mass limit.  As
the DHDA symmetry applies in the heavy quark limit independent of
the number and state of spectator light quarks, it is sufficient
to consider an ordinary heavy baryon, $Qqq$. From DHDA symmetry,
this state is directly related to a doubly heavy tetraquark
state, $\bar{Q} \bar{Q} qq$.  Thus in the heavy quark limit, when
the DHDA symmetry is exact, the existence of heavy baryons
implies the existence of doubly heavy tetraquarks.

The fact that doubly heavy tetraquarks must exist in the heavy
quark limit has been shown previously.  This was done both based
on the simple argument discussed here and in the context of an
illustrative model based on pion exchange \cite{manohar,richard}.
It should be noted that while being in the regime of validity of
DHDA requires the existence of doubly heavy tetraquarks, the
converse is not true: doubly heavy tetraquarks could be formed
via other mechanisms. Nevertheless, the general result is
significant in that the tetraquark has manifestly exotic quantum
numbers in the sense that it cannot be made in a simple quark
model from a quark--anti-quark pair. The observation of exotic
hadrons has been a longstanding  goal of hadronic physics. The
prediction of the existence of an exotic particle directly from
QCD, albeit in a limit of the theory, is of theoretical
importance in that by direct construction QCD is compatible with
exotics.  Other exotic particles, such as a heavy pentaquark,
have also been shown to exist in the heavy quark limit combined
with the large $N_c$ limit \cite{hohler1}.

Let us now turn to more quantitative issues associated with the
excitation spectrum.  As noted in the introduction, the formal
treatment of this problem incorporates NRQCD (for the
interactions between the heavy quarks) and HQET (for the
interactions between the heavy particles and the light degrees of
freedom).  The DHDA symmetry requires each of these effective
theories to be in its domain of validity.  In the heavy quark
limit where both expansions will work, one has
\begin{equation}
M_Q \gg v M_Q \gg v^2 M_Q \gg \Lambda_H \gg
\frac{\Lambda_H^2}{M_Q} \label{scales} \end{equation} where
$\Lambda_H$ is a typical hadronic scale proportional to
$\Lambda_{QCD}$ and $v$, the relative velocity of the heavy
quark, is typically of order $\alpha_s$ and hence depends
logarithmically on the quark mass. It should be noted that the
NRQCD formalism is still valid for $M_Q v^2 \sim \Lambda_H$ as
indicated by ref.~\cite{bvr}. However, none of the analysis in
this work depends on $M_Q v^2$ being larger than $\Lambda_H$, and
hence is consistent with the domain of validity on NRQCD. The
formalism of NRQCD and its associated power counting rules
remains valid for two heavy quarks in the color
$\boldsymbol{\bar{3}}$ in the presence of additional light quark
degrees of freedom and not just for heavy quark--anti-quark
systems in the color singlet in heavy mass limit. This was shown
in ref.~\cite{bvr} and verified in ref.~\cite{fleming}.

It is important to note that these effective theories have
different types of excitations with qualitatively different
scales.  Doubly heavy hadrons (in the formal limit of very large
quark mass) have three characteristic types of excitation:
\renewcommand{\theenumi}{\alph{enumi}}
\begin{enumerate}
\item Excitations of order $\Lambda_H^2/M_Q$ which correspond to the
interaction of the spin of the diquark with the remaining degrees
of freedom in the problem.
\item Excitations of order $\Lambda_H$ which correspond to the
excitations of the light degrees of freedom.
\item Excitations of order $v^2 M_Q$ which correspond to the
internal excitation of the diquark. \end{enumerate} The first two
types of excitations can be understood in terms of HQET while the
third requires NRQCD. The essential point is that as $M_Q
\rightarrow \infty$ the three scales separate cleanly. Since
these excitations all occur at disparate scales, they do not
influence each other.

DHDA symmetry imposes many relations on the various types of
excitations of various doubly heavy hadrons and their associated
singly heavy ones.  To enumerate these, it is useful to have a
naming convention for the various doubly heavy hadrons.  We will
generically call the ground state a doubly heavy baryon with two
$Q$ quarks,  $\Xi_{QQ}$, and the ground state of the tetraquark,
$T_{QQ}$, which are analogous to the heavy (anti-) meson,
$\overline{H}_Q$ ({\it i.e.}, the $\overline{D}$ and
$\overline{B}$ mesons) and heavy Lambdas, $\Lambda$.  We will use
the following convention to indicate various types of hadron
excitations:\\

 ${}^*$ indicates an excitation of type (a);\\

${}^{'}$ indicates an excitation of type (b);\\

${}^\sharp$ indicates an excitation of type (c). \\
\\
In addition, we will indicate the DHDA equivalence between
associating baryons and mesons.

Let us consider the phenomenological consequences of these types
of excitations.  In HQET, the $SU(2)$ heavy spin symmetry causes
states which are only different by a spin flip to have the same
mass.  Excitations of type (a) are the type which will break this
symmetry creating a mass difference between these states. For
example, this will cause a mass difference between the spin-1
$D^*$ meson and the spin-0 $D$ meson.  As this is the leading
term to create the mass splitting, HQET dictates that this
splitting is $O(\Lambda_H^2/M_Q)$ with corrections of
$O(\Lambda_H^3/M_Q^2)$.  Additionally, there are corrections to
this hyperfine splitting due to pNRQCD.  These corrections are
related to the soft gluons that have been integrated out to
construct the diquark potential. The leading corrections
contribute at two loops, as shown in ref.~\cite{bvr}, and are thus
relative $O(\alpha_s^2)$. This implies in a correction to the mass
splitting of $O(\Lambda_H^2 \alpha_s^2/M_Q)$, which is formally
larger than the $O(\Lambda^3/M_Q^2)$ corrections of HQET in the
infinite mass limit. Because $\frac{\Lambda_H}{M_Q}$ is the
smallest scale, these excitations should be the first excitations
above the ground state.

Excitations of type (b) are all other excitations associated with
the light degrees of freedom.  These include orbital excitations
between heavy and light components, as well as excitations within
the light quark degrees of freedom.  Due to the light quark mass,
these excitations are in the non-perturbative regime of QCD, and
can only be characterized by some general hadronic scale,
$\Lambda_H$.  Perturbative corrections to this are, in turn,
meaningless.  Traditional NRQCD has not been applied to systems
with valance light quark degrees of freedom, and thus has ignored
these excitations.  HQET, on the other hand, combines these into
the definitions of heavy fields from the outset, and thereby
neglects them for the rest of the problem.  We see here that the
excitations should be qualitatively the second smallest scale.

Excitations of type (c) are internal diquark excitations.  These
excitations correspond to the excited levels of the color
coulombic potential that binds the diquark.  The binding potential
is $V(r) = - \frac{2}{3} \frac{\alpha_s}{r}$, where the factor of
$\frac{2}{3}$ comes from color considerations.  This leads to
energy levels and energy differences of:
\begin{equation} \label{eq:coloumb}
E_n = - \frac{1}{9} \frac{\alpha_s^2 M_Q}{n^2}; \quad \Delta E =
\frac{1}{12} M_Q \alpha_s^2 = \frac{1}{12} M_Q v^2.
\end{equation}
The last step is justified since at the heavy quark scale,
$\alpha_s(M_Q v) \sim v$.  This verifies that type (c) excitations
are $O(M_Q v^2)$.  This type of excitation should be present in
both the doubly heavy baryon and tetraquark sectors as the light
quark interactions are suppressed since they are $O(\Lambda_H)$.
This leads to mass relations such as:
\begin{equation} \label{eq:sharp}
\Xi_{cc}^\sharp - \Xi_{cc} = T_{cc}^{\Lambda \sharp} -
T_{cc}^\Lambda = T_{cc}^\sharp - T_{cc} = \frac{1}{12} M_Q v^2 +
O(M_Q v^4).
\end{equation}
Since diquark excitations are $O(M_Q v^2)$, these are the largest
excitations discussed here.  The corrections to these relations
can be found by considering the corrections to the color coulombic
potential. In the context of NRQCD, it has been shown by \cite{ms}
that these corrections are $O(M_Q v^4)$ at the heavy quark scale.

In addition to these excitations, DHDA symmetry will relate heavy
mesons, $\bar{Q} q$ states, to doubly heavy baryons, $QQq$
states, and relate heavy baryons, $Qqq$ states, to doubly heavy
tetraquarks, $\bar{Q} \bar{Q} qq$ states, which otherwise have
the same quantum numbers.  Therefore the following relations can
be made:
\begin{equation}
\begin{split}
D &\Leftrightarrow \Xi_{cc}; \\ D^* &\Leftrightarrow \Xi_{cc}^*; \\
\Lambda &\Leftrightarrow T_{cc}^\Lambda; \\
\Sigma, \Sigma^* &\Leftrightarrow T_{cc}, T_{cc}^*, T_{cc}^{**}
\end{split} \label{eq:hda}
\end{equation}
where $D$ and $D^*$ are standard spin-0 and spin-1 D-mesons,
$\Xi_{cc}$ and $\Xi_{cc}^*$ are spin-$\frac{1}{2}$ and
spin-$\frac{3}{2}$ doubly heavy baryons, $\Lambda$ is isospin-0
spin-$\frac{1}{2}$ heavy baryon, $\Sigma$ and $\Sigma^*$ are
isospin-1 spin-$\frac{1}{2}$ and spin-$\frac{3}{2}$ heavy baryons,
$T_{cc}^\Lambda$ is a isospin-0 spin-0 doubly heavy tetraquark,
$T_{cc}$, $T_{cc}^*$, $T_{cc}^{**}$ are isospin-1 spin-0, spin-1,
and spin-2 doubly heavy tetraquarks.

The DHDA symmetry can then be used to relate the mass splittings
\cite{savage}. Equation (\ref{eq:sw}) identifies the corrections
to the mass splitting, but not to the DHDA symmetry itself. DHDA
symmetry relies on the interactions between the heavy diquark and
the light quark(s).  These types of interactions, which are
intrinsically non-perturbative, are not well understood in either
NRQCD or HQET. Therefore, to understand the corrections to the
symmetry, a new power counting scheme that combines the scales of
NRQCD and HQET and is consistent with the other scales in the
problem is necessary to account for these interactions
systematically . At this time, such a system has not been
formulated. Yet we can get a reasonable estimation of the
corrections by considering the effects of the diquark structure
compared with a point-like diquark on the DHDA symmetry. This
consideration is exactly the form factor of the diquark relative
to the scale of the light quark wave function. The form factor
can be calculated by taking the Fourier transform of the square
of the diquark wave function. In the limit of infinite heavy
quarks, the diquark is in a coulombic wave function so the
calculation is straightforward. Assuming that the momentum
transferred is $O(\Lambda_H)$, the form factor can be expanded to
give the leading correction to DHDA symmetry as follows:
\begin{equation} \label{eq:hdacorrect}
F(q) \propto \frac{1}{(1+\frac{a_0^2}{4} q^2)^2} \sim 1 -
\frac{1}{2} a_0^2 q^2 \sim 1 - \frac{1}{2}
\frac{\Lambda_H^2}{M_Q^2 (\frac{2}{3} \alpha_s)^2},
\end{equation} where $a_0$ is the corresponding ``Bohr radius'' of
the coulombic bound state of the diquark. Thus the corrections
due to DHDA are $O(\Lambda_H^2/(M_Q^2 \alpha_s^2))$.  However,
these corrections are formally smaller than the type (a) mass
splitting correction of $O(\alpha_s^2)$. We can translate
Eq.~(\ref{eq:sw}) into the previous notation, and extend the
relations to include the tetraquark splittings to have:
\begin{equation} \label{eq:hdamass}
\begin{split}
\Xi_{cc}^*-\Xi_{cc} &= \frac{3}{4} (D^* - D) + O(\Lambda_H^2 \alpha_s^2/M_Q) \\
T_{cc}^{**}-T_{cc}^* &= \frac{4}{3} (\Sigma^* - \Sigma) + O(\Lambda_H^2 \alpha_s^2/M_Q) \\
T_{cc}^*-T_{cc} &= \frac{2}{3} (\Sigma^* - \Sigma) + O(\Lambda_H^2 \alpha_s^2/M_Q) \\
T_{cc}-T_{cc}^\Lambda &= 2 (\Sigma - \Lambda) + O(\Lambda_H^2
\alpha_s^2/M_Q)
\end{split}
\end{equation}
where all quantities represent the mass of the corresponding
particles.

\begin{figure}[tb]
\includegraphics[scale=.5]{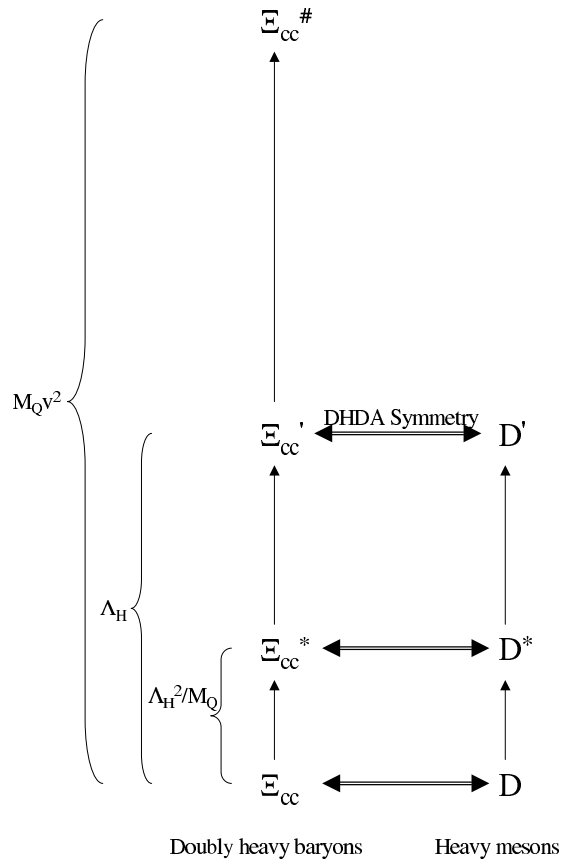}
  \caption{\label{fig:spec1} Hadronic spectrum for doubly heavy baryons related to heavy mesons.}
\end{figure}

\begin{figure}[tb]
\includegraphics[scale=.5]{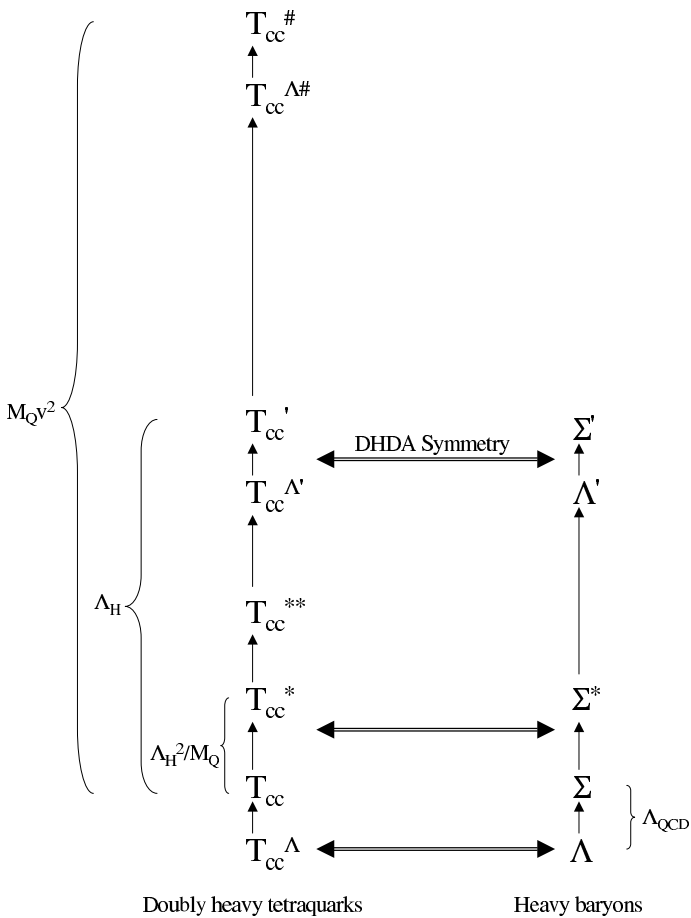}
  \caption{\label{fig:spec2} Hadronic spectrum for doubly heavy tetraquarks related to heavy baryons.}
\end{figure}

To summarize, the doubly heavy baryons and tetraquarks will have
three types of excitations which are distinct in the heavy quark
limit.  From these we can construct the hadronic spectrum for
these particles based upon these excitations and their relative
size to one another.  Additionally, these spectra are related to
the hadronic spectra of heavy mesons and heavy baryons via the
DHDA symmetry.  These spectra are presented in
Figs.~\ref{fig:spec1} and \ref{fig:spec2}.

\section{DHDA symmetry and the physical world}

In the previous section, we have worked solely in the infinite
quark mass limit to determine what the spectrum would look like
in this limit.  We have seen the usefulness of DHDA symmetry in
relating the the spectra of doubly heavy baryons to heavy mesons
and doubly heavy tetraquarks to heavy baryons in this limit. We
would like to use this tool to interpret the corresponding
spectra with a finite massive heavy quark. As the heavy quark
mass is decreased from infinity, we expect that the correction
terms outlined above increase, until at a certain low enough
quark mass, they become as dominant as the leading order
resulting in a break down of the expansion. The discovery of
doubly charmed baryons by the SELEX collaboration provides the
first experimental data to verify the heavy hadronic spectrum
described.  An understanding of the SELEX data can provide an
insight into whether DHDA symmetry persists in the real world.

We can surmise that the SELEX data, along with real world
parameters, could reveal one of three possible insights into the
validity of DHDA symmetry for doubly charmed states. First, upon
examining the data, we could find that the data supports a claim
that the charm mass is heavy enough to be considered in the ideal
large mass limit discussed in the previous section. If this were
the case the spectrum can be easily interpreted in terms of an
approximate DHDA symmetry. Secondly, the opposite could be true,
namely that the SELEX data would be inconsistent with an
approximate DHDA symmetry.  This would indicate that the charm
quark mass is simply too light for the symmetry to be manifest.
The last possibility is perhaps the most interesting, that data
could suggest that charm quark mass is not heavy enough for the
preceding argument to hold in full, but that data would still be
consistent with some aspects of an approximate DHDA symmetry.
This last option is not unreasonable as the DHDA symmetry relies
on the heavy diquark to be view as point-like with respect to the
light degrees of freedom.  The infinite mass limit ensures the
validity of this assumption, but a small-sized diquark might be
achieved even with a relatively modest heavy quark mass.  For
this possibility to be realized dynamics beyond the simple
coulombic interaction must play a central role. To determine
which of these possibilities is most consistent with the SELEX
data, we will examine the size of each of the previously
mentioned excitations, as well as their corrections, and compare
them with experimentally determined parameters from the SELEX
data.

Before doing this, we should note a general word of caution.  The
fact that the excited doubly charmed states were seen only via
their weak decays presents a challenge to {\it any} simple
interpretation of the data.  The problem is that the
electro-magnetic lifetime of the excited states as estimated by
any simple model should be short enough to wash out any detection
of excited states via their weak decay \cite{mehen}. Any simple
interpretation of the SELEX results cannot simultaneously
understand the type of excitation that is observed as well as the
lack of an electro-magnetic decay channel. Therefore our focus
here will be placing limitations on the type of excitation.

The excited state seen by SELEX shown in Fig.~\ref{fig:specxi}
could be interpreted as either a type (a) spin excitation or a
type (c) diquark excitation. Type (b) light quark excitations are
ruled out as they occur on the scale of hadronic physics which is
much larger than the reported excitation. Either interpretation,
as we will discuss, explains aspects of the data, but neither
provides a complete explanation.

\subsection{Scenario I: Spin excitation}

Let us consider the case where the excited states are type (a)
spin excitations. From our discussion of the infinite mass case,
we would expect that even for a finite quark mass, these
excitations would be the lowest lying occurring at
$O(\Lambda_H^2/M_Q)$. According to the SELEX data, the excitation
energy is $78$MeV. With this identification, the DHDA mass
splitting relations, Eq.~(\ref{eq:sw}) and Eq.~(\ref{eq:hdamass})
are satisfied with only a $30\%$ deficiency as has been pointed
out elsewhere \cite{bvr,fleming}. This size of error is also
consistent with the equations' corrections of $O(\Lambda_H^2
\alpha_s^2 /M_Q)$.  This appears to correspond to a success of
DHDA symmetry.

At this point, our predecessors, \cite{bvr, fleming}, have only
verified that Eq.~(\ref{eq:sw}) is satisfied.  This could be
satisfied because DHDA symmetry is the underlying phenomenon or
because of a numerical conspiracy. In order determine between
these two scenarios, one needs to consider the other aspects of
the spectrum and DHDA symmetry. That is, are type (c) excitation
larger than type (a) excitation as expected when a finite quark
mass is considered, and is the spatial extent of the diquark
small enough to consider it point-like?

We will first tackle the former condition. For a diquark bound
solely by color coulombic interactions, the excited state must by
coulombic, and the excitation energy calculated from
Eq.~(\ref{eq:coloumb}) is justified. From Eq.~(\ref{eq:coloumb}),
we can calculate the expected excitation energy of the diquark
for a charm quark mass of $1.15$ GeV and a velocity of $.53$.
This gives an excitation energy of $26.9$ MeV!  This is a clear
sign that the scale separation arising from the color coulombic
interactions, expected for an infinite quark mass, is not present
for the charm quark.

The constrains on DHDA symmetry need to be examined. The key issue
in determining whether DHDA symmetry could hold is the size of
the diquark with regards to the light valance quark(s). This can
be addressed by either looking at the size of the diquark to
determine if it is nearly point-like, or to determine the size of
corrections of DHDA symmetry as shown in
Eq.~(\ref{eq:hdacorrect}).  The size of the diquark can be
characterized by the RMS radius of the state. For coulombic wave
functions, the size of the diquark in the ground state is $1.64$
fm. Clearly this is not point-like on the scale of hadronic
physics. The large size of the ground state of the diquark also
indicates that the excited state would be even larger. Such a
spatially large excited state suggests that the excited state
should extend beyond the color coulombic potential.  This
invalidates the previous calculation, while emphasizing the
absurdity of assuming that the diquark is bound deeply by the
color coulombic interaction. Moreover, this further indicates that
the diquark must be under the influence of interactions in
addition the the color coulombic potential. Additionally, the
corrections to DHDA symmetry should be small compared with $1$ if
the approximation is used. For the values for the charm quark, the
correction can be calculated, from Eq.~(\ref{eq:hdacorrect}) to
be $3.02 \frac{\Lambda_H^2}{\text{GeV}^2}$, which for a typical
hadronic scale of $\Lambda_H \sim 1$ GeV is not much smaller than
1.  Thus both indicators show that the real world charm quark is
not heavy enough to justify the point-like nature of the doubly
heavy diquark which is necessary for the DHDA symmetry.

It should be noted that the bottom quark has a mass marginally
large enough to approach the infinite mass limit scaling.  The
type (c) excitation is $32.8$ MeV, with the type (a) excitation
being $34.3$ MeV calculated from the B-meson mass splitting.
Additionally, the characteristic size is $0.79$ fm, and the
correction to the DHDA symmetry is $.69
\frac{\Lambda_H^2}{\text{GeV}^2}$. All of these numbers show that
for the bottom quark the scale hierarchy is as expected and
corrections are relatively small, even if the scale separation is
not complete. However, presently doubly bottom baryons have not
been observed experimentally.

We have shown that a naive approach to DHDA symmetry results in
the conclusion that the charm quark is by no means heavy enough
to believe that this symmetry is manifest in the real world, at
least if it is to arise due to color coulombic interactions. In
other words, the relatively small charm quark mass causes the
corrections to the infinite massive limit to become large enough
to question the expansion for the excited states. However, this
does not completely rule out the possibility that DHDA could hold
approximately and that these excitations are type (a). The color
coulombic interactions are not the only interactions that the
charm quarks could experience as part of a diquark or a doubly
heavy baryon. Since the charm quarks are not heavy enough to fall
into the color coulombic region, it is reasonable to surmise that
these other non-perturbative interactions could conspire in such
a manner that would facilitate an approximate DHDA symmetry.
However, these additional non-perturbative interactions are not
systematically included in NRQCD.  Therefore, in order to
describe this system, a new expansion that combines the
perturbative and non-perturbative scales of NRQCD and HQET in a
systematic manner is needed. At present, such an expansion has not
yet been formulated. Nevertheless, by examining the properties of
the interactions needed to maintain DHDA symmetry, a general
picture of the new theory could be made.

Before proceeding with a discussion of the conditions that DHDA
symmetry imposes on additional non-perturbative interactions, an
additional comment on the color coulombic potential is needed.
First, when we worked in the large mass limit, we were required
to be in the regime of $M_Q v^2 \gg \Lambda_H$. However, with a
finite massive quark this condition could be weakened to include
$M_Q v^2 \sim \Lambda_H$. Under this condition, the type(b) and
type (c) excitation may mix since they are at the same energy
scale. Nevertheless, the key issue here is whether type (a) and
type (c) excitations separate. The possible inclusion of type (b)
excitations with type (c) does not effect whether they are
separated from type (a), and hence do not effect the results
discussed here. Secondly, the color coulombic potential is only
the leading order term in NRQCD; sub-leading terms might need to
be included when a finite massive quark is considered. However,
since we have seen a need for a new expansion that includes the
mixture of perturbative and non-perturbative effects, it is not
clear whether the sub-leading terms suggested by NRQCD are the
only sub-leading terms in the combined expansion. In both of these
cases though, additional interactions beyond the simple color
coulombic potential are included.  It is not unreasonable that
these, just like the ones hypothetically postulated above, would
conspire so that the DHDA symmetry would be manifest in an
approximate manner in the real world. Again, a description of the
conditions to obtain an approximate DHDA symmetry will provide
insight into these additional interactions whether they are NRQCD
based or well beyond the scope of NRQCD and HQET.

There are two key places where the analysis based on the color
coulombic potential fails to give rise to the DHDA symmetry with
real world parameters. The assessment of these failures will
provide conditions on the additional interactions to reestablish
DHDA symmetry. The first is the characteristic size of the
diquark. We have already shown that for the coulombic potential,
the size of the diquark is large enough not to be considered even
remotely point-like from the point of view of hadronic dynamics.
Secondly, the hierarchy of scales used to derive  the result
breaks down badly.  Additional dynamics beyond color coulombic
would need to create a diquark with a size much smaller than the
characteristic hadronic size and to re-establish the spin
excitations as the lowest lying excitations as we originally
assumed.

An examination of the restrictions placed on the characteristic
size of the diquark reveals the following. The characteristic
size of the diquark, which we will denote as $L$, must be smaller
than the size in a coulombic potential, denoted $L_c$, and it must
be small enough to allow the DHDA corrections to be small. The
correction term of Eq.~(\ref{eq:hdacorrect}) can be rewritten in
terms of this characteristic size as $\frac{1}{6} L^2
\Lambda_H^2$.  Thus for the correction to be small $L \ll
\sqrt{6}/\Lambda_H \equiv L_{DHDA}$. $L_c$ must be larger than
$L_{DHDA}$ since $L_c$ already violates DHDA symmetry and thus
cannot be smaller than $L_{DHDA}$. Therefore, in order for the
diquark to be considered point-like, both
\begin{equation}\label{eq:ldhda} L\ll L_c \quad \text{and} \quad
L \ll L_{DHDA}\end{equation}must be simultaneously satisfied. In
order to insure this, in terms of size, $L_{DHDA}$ could be much
smaller than $L_c$, or $L_{DHDA}$ could be of comparable size to
$L_c$. Consider the former possibility. $L_{DHDA} \ll L_c$ is
equivalent to $\frac{\sqrt{6}}{\Lambda_H} \ll \frac{3}{2 M_Q
\alpha_s}$. This implies that $M_Q \alpha_s \ll .6 \Lambda_H$.
This relationship is never satisfied since $\alpha_s \sim
1/\ln(M_Q)$ and $M_Q \gg \Lambda_H$.  Thus for DHDA symmetry to
occur the latter condition must hold. It gives: $L_{DHDA} \sim
L_c$ implying $M_Q \alpha_s \sim .6 \Lambda_H$. As $\alpha_s$ at
the charmed quark mass scale {\it is} around $.6$, this relation
can only be satisfied if $M_Q \sim \Lambda_H$. It should be noted
however, that an interaction that provides a characteristic size
of the diquark which is consistent with Eq.~(\ref{eq:ldhda}) is
possible. For the purposes of our discussion here, we needed to
show that at least one kinematic region was possible, and the
region where $M_Q \sim \Lambda_H$ satisfies these conditions even
though it should not be unique.

A couple of comments should be made about this condition.  The
first is that naively appears not to occur even for the charm
quark case. If one takes $\Lambda_H$ to be of the scale of
$\Lambda_{QCD}$ it seems to be much smaller than $M_c$. However,
we should note that the $\sim$ indicates ``of the same scale as''
under the assumption that the coefficients which arise in the
expansion are ``natural'' {\it i.e.} of order unity. If the
dynamics are such that some of the coefficients multiplying
$\Lambda_H$ are anomalously large, the condition $M_Q \sim
\Lambda_H$ could hold effectively.    The second key point is
simply that if this does occur the system is clearly beyond the
perturbative regime.  It should also be noted that that this
should {\it not} be seen as a generic condition invalidating
NRQCD.  Rather it implies that {\it for this particular system}
the expansion has broken down.  There is non-trivial evidence that
this is in fact the case; namely if one assumes that the
expansion is working one gets inconsistent results. The central
question addressed here is not whether the expansion has broken
down, rather it is whether one can still have a small diquark
even if the expansion has broken down.  If it  indeed is the case
that the condition $M_Q \sim \Lambda_H$ is effectively met, then
there is a possible characteristic size of the charmed diquark,
for which DHDA symmetry could be valid. This region is simply a
size that is much smaller than the length associated with the
coulombic potential and smaller than the typical hadronic size.

Thus far we have identified a possible kinetimatic region for
which approximate DHDA symmetry may be possible. However, to test
whether this can occur in practice, we need to see whether
plausible dynamics can drive the system into such a regime.  We
do this by considering a ``reasonable'' dynamical model for the
interaction between the heavy quarks. This model is not intended
to be an accurate description of hadronic physics.  The goal is
simply to see whether a simple model with natural scales can put
the system in the regime where DHDA symmetry emerges at least
approximately. The existence of a model which does this shows
that an approximate DHDA symmetry could be present in charm
physics despite the fact that NRQCD in the coulombic regime plus
HQET alone do not give rise to an approximate DHDA symmetry with
the real world charm quark mass.

To illustrate the kind of model which brings us into this regime,
we consider a linear confining potential with a string tension of
$1 \frac{\text{GeV}}{\text{fm}}$.  Such a potential, with the same
string tension, can be used to get a reasonable description of the
$J/\Psi$ \cite{eichten}.  One might not believe that such a model
is applicable at all distances, to which we will attempt to apply.
Indeed, one may reasonably question whether any two-body potential
description is sensible.  Nevertheless the scales of the model
are at least instructive. Any confining potential that can be
introduced will cause the characteristic size of the diquark to
be reduced, thus the conditions on the diquark size may be
satisfied. Specifically for the linear confining potential above
binding charmed quarks, the characteristic length is $0.5$ fm.
This is substantially smaller than the coulombic wave function and
might be small enough so that approximate DHDA might emerge.
Moreover, the small size of the bound state helps to justify the
two-body potential description {\it a posteriori}; the effects of
the light quark between the heavy ones should be suppressed due to
the small size. Unfortunately, this calculation is not part of a
systematic calculation, and it is not immediately clear how to
reliably estimate the size of the correction to the leading order
DHDA estimate for the splitting.

Calculations of the energy spectrum of coulombic plus linear
confining potentials in this channel reveals that the radial
excitation energy is $630$ MeV, far above the $100$ MeV energy
associated with expected spin excitations. Thus this linear
confining potential satisfies both of the conditions needed to
believe that an approximate DHDA symmetry could be realized for
charmed quarks.

We have thus found a region where an approximate DHDA symmetry
could be realized approximately and the lowest lying excitations
are type (a) spin excitations.  The color coulombic interactions
cannot be the only relevant interactions that the heavy quarks
experience (as is assumed in the heavy quark mass limit).  Of
course the question of whether the dynamics as such is realized
in nature, remains an open question.

Even though we have provided a consistent argument for the
observed excited states to be spin excitation, there remains a
phenomenological issue with the parity of the excited state.  Type
(a) excitations do not change the parity of the excited state
relative to the ground state. Ground state baryons  have positive
parity, thus the spin excited state should also have positive
parity. Experimentally, the parity of the excited states has not
been determined.  The SELEX collaboration have argued that the
orbital angular momentum of the ground state is consistent with
$L=0$ (positive parity), while the excited state is consistent
with $L>0$ (either positive or negative parity). Furthermore,
SELEX observed an orbital excited state $\Xi_{cc}(3780)$ which has
negative parity and decays via pion emission to $\Xi_{cc}(3520)$
suggesting that this state could have negative parity.  If this
parity assignment holds, the interpretation that the excited
states were spin excitations, made here and in
refs.~\cite{bvr,fleming}, would be ruled out.

\subsection{Scenario II: Diquark Excitation}

Now let us consider the case where the excitation is interpreted
as a type (c) diquark excitation.  Type (c) excitations could
result in a parity flip from the ground state.  This would resolve
the parity problem found with the spin excitation
interpretation.  As we will discuss below, if this scenario is
correct we are almost certainly outside the regime of validity of
DHDA as well as outside the regime of validity of NRQCD.
Moreover, it is likely to be very difficult to make such a
scenario work phenomenologically.

In order for diquark excitations to be smaller than the spin
excitations, there must be a break down of the heavy quark mass
limit;  the system must reside in a non-perturbative regime.
Therefore, as with the previous case, the diquark can be under
the influence of non-perturbative interactions beyond the color
coulombic interactions.  We showed that if these additional
confining interactions maintained an approximate DHDA symmetry,
the diquark excitations were much larger than the observed $78$
MeV excitation.  Again we can illustrate this with a linearly
rising potential between the heavy quarks.  In order for the
diquark excitations to be comparable to the observed splitting,
the linear confining interactions must have a string constant of
$\sim 50 \frac{\text{MeV}}{\text{fm}}$, which is very small
compared to the natural scales in the problem. The small size of
the linear confining interactions results in the diquark having a
larger size, and makes the assumptions that it is point-like even
less believable.  For the string constant of 50
$\frac{\text{MeV}}{\text{fm}}$ considered here, the ground state
of the diquark has an RMS radius of $1.2$ fm, and the first
excited state has an RMS radius of $5.8$ fm.  These numbers are
extremely large compared to typical hadronic sizes.

The preceding model calculation suggests that if the excitation
were the excitation of the diquark, the DHDA symmetry cannot be
valid even approximately. It also raises a fundamental issue of
self-consistency.  A large spatially extent diquark allows the
light quark to come between the two heavy quarks allowing for
three-body interactions to play a significant dynamical role.  To
the extent that this occurs, it is meaningless as a
phenomenological matter to separate the diquark excitation from
excitations of the entire system.  Thus, excitations of type (b)
and (c) would strongly mix and the entire structure of scale
separation would break down.

We note that this analysis was based on a very simple and not
terribly plausible model. However, it does incorporate the
natural scales of the problem and shows that the excited state
wave functions are {\it much} too large to be taken seriously. It
is also clear that it would be very hard to construct any
potential model which restricts the diquark size to be much less
than a fermi while having an excitation energy of $78$ MeV. To
illustrate this point, we can consider a harmonic confining
potential instead of the linear potential. We would expect that
this potential would confine the excited state and reduce its
size more than the linear potential. Calculating the size of the
diquark under these conditions for an excitation energy of $78$
MeV results in a ground state RMS radius of $1.3$ fm and an
excited state RMS radius of $2.5$ fm. Even though the diquark
size is smaller, it is still very large in terms of hadronic
physics. Furthermore, if we were able to drive the size of the
excited state to a reasonable hadronic size, say, $1$ fm, the
ground state would be even smaller. Such a small ground state
size is then consistent with spin excitations discussed
previously. It thus seems difficult for this scenario to be
correct.

Together these two scenarios make it very difficult to understand
the data in a simple way. If the parity of the is the parity of
the states is correctly interpreted by SELEX, there does not
appear to be any simple phenomoenologically reasonable
interactions yielding either small diquark excitations or DHDA
symmetry. However, if we set the parity designation aside,
scenario I, the assignment of type (a) spin excitations, seems to
be the more plausible interpretation.

\section{Conclusion}

The SELEX data on doubly heavy baryons is very difficult to
interpret.  As noted in the introduction, the fact the excited
states were detectable through their weak decays when there were
open channels for electromagnetic decays is very problematic;
normally one would expect these to dilute the strength to the
point that the states would be very difficult to see. Despite
this problem, we have attempted to understand the SELEX states in
the context of DHDA symmetry. We have shown that the data is not
consistent with the heavy quark mass limit, but this does not rule
out an approximate DHDA symmetry.  This could emerge if diquark
interactions beyond color coulombic interactions are considered.
As such, a new systematic expansion, which is a hybrid of HQET
and NRQCD but outside the domain of the color coulombic, would be
the most appropriate to describe the physics of doubly heavy
baryons. Such an expansion could help in the understanding of the
SELEX observations. At the same time it is critical to add to our
understanding of the experimental situation.  In particular, it is
essential that the observed states be confirmed in other
experiments; that the parity of the states are pinned down; and
the accurate measurements of electromagnetic transitions are made.
These measurements are critical in understanding the doubly charm
baryon spectrum as well as the validity of DHDA symmetry in charm
physics.

{\it Acknowledgments.}  T.D.C.\ and P.M.H.\ were supported by the
D.O.E.\ through grant DE-FGO2-93ER-40762. The authors would also
like to thank Nora Brambilla and Thomas Mehen for their insightful
comments.

\end{document}